\documentclass[10pt,twocolumn,english,aps,manuscript,aps,showpacs,showkeys,superscriptaddress,citeautoscript]{revtex4}
\usepackage[T1]{fontenc}
\usepackage[latin9]{inputenc}
\usepackage{textcomp}
\usepackage{amsmath}
\usepackage{amssymb}
\usepackage{graphicx}
\usepackage{esint}

\makeatletter
\@ifundefined{textcolor}{}
{%
 \definecolor{BLACK}{gray}{0}
 \definecolor{WHITE}{gray}{1}
 \definecolor{RED}{rgb}{1,0,0}
 \definecolor{GREEN}{rgb}{0,1,0}
 \definecolor{BLUE}{rgb}{0,0,1}
 \definecolor{CYAN}{cmyk}{1,0,0,0}
 \definecolor{MAGENTA}{cmyk}{0,1,0,0}
 \definecolor{YELLOW}{cmyk}{0,0,1,0}
}

\usepackage{babel}
\include{shadowtext}

\makeatother

\usepackage{babel}
\begin{document}

\title{Non-conserved magnetization operator and `fire-and-ice' ground states
in the Ising-Heisenberg diamond chain.}

\author{Jordana Torrico}

\affiliation{Instituto de Fisica, Universidade Federal de Alagoas, 57072-970,
Maceió, AL, Brazil}

\author{Vadim Ohanyan}

\affiliation{Laboratory of Theoretical Physics, Yerevan State University, Alex
Manoogian 1, 0025 Yerevan, Armenia}

\author{Onofre Rojas}

\affiliation{Departamento de Fisica, Universidade Federal de Lavras, CP 3037,
37200000, Lavras, MG, Brazil}

\begin{abstract}
We consider the diamond chain with S=1/2 XYZ vertical dimers which interact with the intermediate sites via the interaction of the Ising type. We also suppose all four spins form the diamond-shaped plaquette to have different g-factors. The non-uniform g-factors within the quantum spin dimer as well as the XY-anisotropy of the exchange interaction lead to the non-conserving magnetization for the chain. We analyze the effects of non-conserving magnetization as well as the effects of the appearance of negative g-factors among the spins from the unit cell. A number of unusual frustrated states for ferromagnetic couplings and g-factors with non-uniform signs are found out. These frustrated states generalize the "half-fire-half-ice" state introduced in Ref. [\onlinecite{yin15}]. The corresponding zero-temperature ground state phase diagrams are presented.
\end{abstract}


\maketitle

\section{Introduction}

In the last decade, intensive investigations have been focused on
the effects of magnetic anisotropy in metal complexes and adatoms. The anisotropy
arises due to the interplay of the spin-orbit coupling on the
magnetic ion sites and the crystal field from neighboring atoms and ligands \cite{jomie,carlin,kahn}. This
phenomenon can affect the magnetothermal properties of the system essentially \cite{oha15,tor16}. One of the most unusual features of these joint interactions is the negative Land\'{e} g-factor which occurs in some complexes \cite{ungur,fecu,yin15}. The appearance of the negative and positive g-factors in the same system leads to a series of peculiar features even in the simplest case of Ising chain with alternating g-factors. It was demonstrated in Ref. [\onlinecite{yin15}] that the novel frustration can be arisen in ferromagnets with non-uniform g-factors with different signs. It was also argued in the paper that the aforementioned novel frustrated state, which has been given the name "half-fire-half-ice" by the authors can be realized in copper-iridium oxides such as Sr$_3$CuIrO$_6$ \cite{Ir1, Ir2}. Also, the magnetic centers in some compounds
of the transition-metal ions with unquenched angular momentum and relatively
strong spin-orbit coupling could posses rather large Land\'{e} g factors, essentially different from the corresponding g-factors for free ion.
One can mention, for instance, Fe$^{3+}$ ion with a Land\'{e} g factor $g\approx 2.8$, as
well as Co$^{2+}$ ion with $g\approx 6.0$ \cite{jomie,carlin,kahn}.

Large anisotropy can be obtained combining almost isotropic transition-metal
ion with highly anisotropic rare-earth ions increasing the difference
of the Landé g factors in oligonuclear complexes. In it known that the Dy$^{3+}$ ion
has roughly $g\approx20$. A series of magnets compounds with this ion have been recently investigated revealing some intriguing properties \cite{Dy1,Dy2,bel14}.  These unusual large g factors must correlate
with a strong anisotropy in the exchange interaction as well \cite{jomie,mironov,chibotaru}.
One can mention the heterodinuclear Cr$^{3+}$-Yb$^{3+}$ \cite{mironov} complex as an example of the molecular magnet with
 highly anisotropic exchange interaction in $z$ direction.

A recent investigation of the magnetism of a Co$_{5}$ complex brings evidence
of negative g factors for some Co$^{2+}$
ions \cite{klo09}. Surely, this study stimulates a deeper understanding
of the origin of negative g factors and their implications for magnetic
properties of some compounds. The inversion of the sign of the g factors can occur in the molecular magnets as well as in the single chain magnet and other materials \cite{yin15}. For instance, in Ref. [\onlinecite{fecu}] combining ligand field and density functional theory (DFT) analysis of the magnetic anisotropy in cyanide-bridged single-molecule magnets (oligonuclear complexes, Fe$^{\mbox{III}}$---CN---M$^{\mbox{II}}$ (M=Cu, Ni)) has been performed. Particularly,  it was found that the g-factor of the Fe$^{3+}$ ion is isotropic and negative, $g_{\mathrm{Fe}}=-1.72$, while for the and Cu$^{2+}$ ion it is positive and has small axial anisotropy, $g_{\mathrm{Cu}_x}=g_{\mathrm{Cu}_y}=2.18$, $g_{\mathrm{Cu}_z}=2$. It was also shown recently using the \textit{Ab initio} calculation that the product of the diagonal components of the Lande g factors satisfy for some lanthanide and transition metal complexes $g_{x}g_{y}g_{z}<0$. It is worth mentioning that the negative sign of the product of Land\'{e} g-factor components has been known for some transitional metals and lanthanide complexes since 60s \cite{ggg1,ggg2,ggg3}.


Moreover, there are compounds of single-chain magnet (SCM) type, for example
$[{(\mathrm{CuL})_{2}\mathrm{Dy}}{\mathrm{Mo(CN)}_{8}}]\cdot2\mathrm{CH_{3}CN\cdot H_{2}O}$ \cite{Dy1,Dy2}
which are an interesting magnetic material exhibiting different Landé g-factors for different magnetic ions
and describing within the Ising-Heisenberg spin chain
model. These models, in contrast to the Ising-Heisenberg models with uniform g-factors demonstrate zero temperature magnetization curve with an unusual non-plateau behavior within the same eigenstate \cite{oha15, tor16, bel14, str05}.
 The theoretical model of the aforementioned compound can be solved exactly by means of the generalized classical transfer matrix method \cite{bel14}. The models of the Ising-Heisenberg type imply the lattice consisting of small quantum spin clusters interacting with each other through the intermediate Ising spin \cite{tor16, bel14, str05, str03, can06, val08, can09, ant09, bel10, roj11a, oha12, bel13, gal13, ver13, gal14, ana14, tor14, qi14, lis15, abg15, gal15, gao15, lis16, hov16, ana17, rod17}. Therefore, the eigenstates of the whole system are direct products of the eigenstates for the quantum spin clusters. The zero-temperature magnetization curve of such models usually contains the regions corresponding to certain eigenstates with the sharp transitions between them. These regions are horizontal (magnetization plateaus) in case if the magnetic moment is a good quantum number and each eigenstate possesses fixed value of it. This is the case for conserving magentization operator. However, for the different g-factors for different spins within the same cluster the magnetizaiton operator does not commute with the Hamiltonian. As a result the magnetic moment is not a good quantum number and the magnitude of magnetization could vary within the same eigenstate under the change of the magnetic field. Thus, the deviations form the horizoantal line is occur in the magnetization curve (quasi-plateaus) \cite{oha15, tor16, bel14, str05}. However, the deviation of the magnetization curve parts from the horizontal line due to difference in g-factors of the quantum spin from the three-spin linear cluster in the $[{(\mathrm{CuL})_{2}\mathrm{Dy}}{\mathrm{Mo(CN)}_{8}}]\cdot2\mathrm{CH_{3}CN\cdot H_{2}O}$ SCM is merely visible by eyes, as the difference of the values of g-factors is rather small \cite{bel14}. Almost the same effect has been observed but even quantitatively less pronounced in the approximate model of the SCM, the F-F-AF-AF spin chain compound$\mathrm{Cu}(3\mathrm{-chloropyridine})_{2}(\mathrm{N}_{3})_{2}$
\cite{str05}.

    In the past decades, a so-called diamond chain magnetic structure and its variants have been
intensively studied. Since the experimental discovery that the Cu$^{2+}$ ion in the well-known mineral azurite, $\mathrm{Cu_{3}(CO_{3})_{2}(OH)_{2}}$, are arranged along the $b$-plane in a diamond chain manner and that the interchain coupling is small enough \cite{kikuchi03}, the issue has been receiving permanent attention form the theoreticians and experimentalists \cite{azu4, heschke, hid17, mor17}. Due to its symmetric properties and relative simplicity the diamond chain is also the most popular one-dimensional structure for the theoretical research in the field of the Ising-Heisenberg spin lattices. Various physical effects and issues have been considered in the context of the corresponding model on the diamond chain or its modification, magnetization plateaus and zero-temperature phase diagrams, higher spin, mixed spins, four-spin interaction, magnetocaloric effect, entanglement and quantum state transfer, just to mention few of them   \cite{oha15, tor16, can06, can09, roj11a, bel13, gal13, gal14, ana14, tor14,qi14, lis15, abg15, gal15, gao15, lis16, hov16}.

 In the present paper we consider the S=1/2 Ising-Heisenberg model on the diamond chain with non-conserved magtnetization due to non-uniform g-factors as well as due to XY-anisotropy. We describe the eigenstates of the chain for the case of four different g-factors. Particularly, we are interested in the zero-temperature effect induced by the appearance of the negative g-factors(s). As a further development of the ideas of the Ref. [\onlinecite{yin15}] we preset the detailed description of the "fire-and-ice" configuration which in our case are more divers. We analyze the Ising case as well as the whole Ising-Heisenberg model.

The paper is organized as follows. In Sec. 2 we present
the model under consideration and make a general statements about the non-commutativity of the magnetization
and the Hamiltonian, its origin and basis consequences.  In Sec.3 we describe in details the ground states
of the model and its Ising limit. In sec. 4 we study the effect of the negative g-factor for the part of the spins from the unit cell. We found various frustrated states of the "fire-and-ice" type introduced in Ref. [\onlinecite{yin15}]. The Sec. 5 contains a conclusion.

\section{The model}

\begin{figure}[t]
 \includegraphics[width=1\columnwidth]{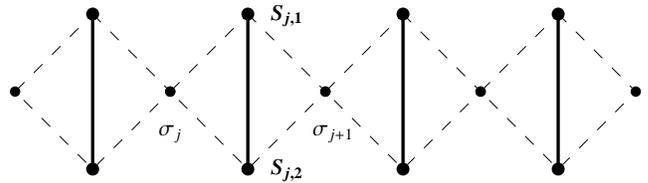} \caption{\label{fig1} The Ising-Heisenberg diamond-chain. Solid lines represent
the quantum interactions, while the dashed one stand for the interaction
involving only $z$-components of the spins. Here we also consider
the $g$-factors of the quantum spins $\mathbf{S}_{j,1}$ and $\mathbf{S}_{j,2}$
to be $g_{1}$ and $g_{2}$ respectively, and $\sigma_{j}$ and $\sigma_{j+1}$
to be $g_{3}$ and $g_{4}$.}
\end{figure}

Let us consider the $S=1/2$ $XYZ$-Ising diamond chain describing by the following
Hamiltonian (See Fig. \ref{fig1})
\begin{alignat}{1}
\mathcal{H}= & \sum_{j=1}^{N}\left(\mathcal{H}_{j}-Bg_{j}\sigma_{j}\right),\label{Hamdc1}
\end{alignat}
where $\mathcal{H}_{j}$ is given by

\begin{alignat}{1}
\mathcal{H}_{j}= & J\!\left\{ \left(1\!+\!\gamma\right)S_{j,1}^{x}S_{j,2}^{x}\!+\!\left(1\!-\!\gamma\right)S_{j,1}^{y}S_{j,2}^{y}\right\} \nonumber \\
 & +\Delta S_{j,1}^{z}S_{j,2}^{z}+K(S_{j,1}^{z}\!+\!S_{j,2}^{z})(\sigma_{j}\!+\!\sigma_{j+1})\nonumber \\
 & -B(g_{1}S_{j,1}^{z}\!+\!g_{2}S_{j,2}^{z}),\label{eq:H_j}
\end{alignat}
the g-factors of the Ising intermediate spins, $\sigma_j$ are supposed to be alternating,
\begin{equation}
g_{j}=\left\{ \begin{array}{cc}
g_{3}, & j\;\mbox{is odd}\\
g_{4}, & j\;\mbox{is even}.
\end{array}\right.\label{g34}
\end{equation}
Thus, the diamond-chain is composed of the vertical $S=1/2$ $XYZ$-dimers
with quantum spin operators $\mathbf{S}_{j,1}$ and $\mathbf{S}_{j,2}$.
These dimers are alternating with Ising spins $\sigma_{j}$ taking
$\pm1/2$ values. The Ising spins interact with the $z$-component
of their left and right neighboring $\mathbf{S}$-operator with exchange
interaction $K$. The quantum spins belonging to the same dimer are
also supposed to have different Landé g-factors, denoted by $g_{1}$
and $g_{2}$. Therefore, the Hamiltonian of the whole system is the sum
of the mutually commutative block Hamiltonians $\mathcal{H}_{j}$.
The important feature of the Hamiltonian $\mathcal{H}$ is its the non-commutativity
with the magnetization operator,
\begin{eqnarray}
\mathcal{M}^{z} & = & \frac{1}{N}\sum_{j=1}^{N}\left(g_{1}S_{j,1}^{z}+g_{2}S_{j,2}^{z}\right)\nonumber \\
 &  & +\frac{1}{N}\sum_{j=1}^{N/2}\left(g_{3}\sigma_{2j-1}+g_{4}\sigma_{2j}\right),
\end{eqnarray}
\begin{eqnarray}
\left[\mathcal{H},\;\mathcal{M}^{z}\right]\neq0.\label{comm1}
\end{eqnarray}
The origin of these non-commutativity is the difference in g-factors
for the quantum spins and $XY$-anisotropy,
\begin{eqnarray}
&&\left[\mathcal{H}_{j},g_{1}S_{j,1}^{z}\negthinspace+\negthinspace g_{2}S_{j,2}^{z}\right]= -i\gamma\left(g_{1}+g_{2}\right)\negthinspace\left(S_{j,1}^{x}S_{j,2}^{y}+S_{j,1}^{y}S_{j,2}^{x}\right)\nonumber \\
 && +iJ\left(g_{1}-g_{2}\right)\negthinspace\left(S_{j,1}^{x}S_{j,2}^{y}-S_{j,1}^{y}S_{j,2}^{x}\right).
\end{eqnarray}

As one can see, there are two sources of the non-commutativity, the
$XY$-anisotropy $\gamma$ and difference of the g-factors $(g_{1}-g_{2})$ \cite{oha15}.
This non-commutativity leads to a non-linear magnetic field dependence
of the spectrum of the model and to the phenomena of quasi-plateau \cite{oha15,bel14}.
The quasi-plateau actually means the eigenstate with an explicit magnetic
field dependance, even at zero temperature. The part of the magnetization
curve corresponding to the eigenstate with an explicit magnetic field
dependance demonstrates the monotonous grow of the magnetization with
the increasing the magnetic field magnitude, instead of being constant
(plateau) what takes the place in a conventional case when the finite
spin cluster has conserving magnetization operator. Non-commutativity
of the magnetization operator and Hamiltonian leads to another unusual
phenomena, the reentrant transitions due to non-linear magnetic field
dependence of the spectrum. The sequence of the quantum phase transitions
at zero temperature with the monotonous changing of the magnetic field
for a finite spin cluster is determined by a level crossing. For the
linear in magnetic field spectrum any two levels can have no more
than one crossing and thus each eigenstate can appear only once in
the magnetization curve. In case of non-linear spectrum two levels
can have more than one crossing which can lead to a multiple appearance
of the same ground state in the magnetization curve.

As the lattice has six spins in the translational invariant unit cell,
two $\sigma$ spins and two vertical dimers, the total saturation
magnetization per unit cell (note that $N$ is the number of block
which is supposed to be even, while the number of the unit cell with
six spins is $N/2$) is
\begin{equation}
M_{sat}=\left(g_{1}+g_{2}+\tfrac{1}{2}\left(g_{3}+g_{4}\right)\right).\label{mag0_dc}
\end{equation}

\section{Ground states}

The eigenstates of the chain are composed of a direct product of the
eigenstates of each block. The Ising interaction between the vertical
dimers makes the propagation of any type of spin excitation from block
to block impossible. That is why, we can describe all possible ground
states of the system exactly in term of few configuration. However,
the Hamiltonian breaks the translational symmetry of the diamond-chain
by the doubling of the block leading to the six-spin unit cell. When $g_3$=$g_4$ the unit cell coincides with the three-site triangular block of the diamond chain.
\subsection{Quantum dimer eigenstates}

Let us start with the description of the four eigenstates of the isolated quantum
spin dimer (Eq. \eqref{eq:H_j}), which are the building blocks for the
construction of the ground states for the whole chain. After diagonalization of
the block Hamiltonian \eqref{eq:H_j}, we obtain four eigenvalues. The first
couple of eigenvalues $\varepsilon_{1,2}$ can be expressed as follows:
\begin{equation}
\varepsilon_{1,2}\left(\sigma_{j},\sigma_{j+1}\right)=-\tfrac{\Delta}{4}\pm G,
\end{equation}
where $G=\tfrac{\sqrt{B^{2}(g_{1}-g_{2})^{2}+J^{2}}}{2}$. The corresponding
eigenstates are independent on the value of the neighboring $\sigma$-spins:
\begin{alignat}{1}
|\Psi_{1,2}\rangle= & \frac{\left(|\uparrow\downarrow\rangle+c_{\pm}|\downarrow\uparrow\rangle\right)}{\sqrt{1+c_{\pm}^{2}}},\label{psi12}
\end{alignat}
with $c_{\pm}=\frac{B(g_{1}-g_{2})\pm2G}{J}.$

This eigenstates in the limit of uniform g-factors transform to the
singlet state and $S^{z}=0$ component of the triplet state. That
is why the vertical dimer decouples from its neighborhood.

The second set of eigenvalues $\varepsilon_{3,4}$ of the Hamiltonian
\eqref{eq:H_j} are
\begin{equation}
\varepsilon_{3,4}\left(\sigma_{j},\sigma_{j+1}\right)=\tfrac{\Delta}{4}\pm F_{\sigma_{j},\sigma_{j+1}},
\end{equation}
 where $F_{\sigma_{j},\sigma_{j+1}}=\tfrac{\sqrt{\left[B(g_{1}+g_{2})-2K(\sigma_{j}+\sigma_{j+1})\right]^{2}+J^{2}\gamma^{2}}}{2}.$
The two eigenstates associated with the $\varepsilon_{3,4}$ eigenvalues
are dependent on their left and right $\sigma$ spins:
\begin{alignat}{1}
|\Psi_{3,4}\rangle= & \frac{\left(|\uparrow\uparrow\rangle+b_{\sigma_{j},\sigma_{j+1}}^{\pm}|\downarrow\downarrow\rangle\right)}{\sqrt{1+(b_{\sigma_{j},\sigma_{j+1}}^{\pm})^{2}}},\label{psi34}
\end{alignat}
with $b_{\sigma_{j},\sigma_{j+1}}^{\pm}=\frac{B(g_{1}\!+\!g_{2})\!-\!2K(\sigma_{j}\!+\!\sigma_{j+1})\!\pm\!2F_{\sigma_{j},\sigma_{j+1}}}{J\gamma}$.

Thus, here we have three different eigenstates for a vertical quantum
dimer depending on the configuration of neighboring $\sigma$ spins:
$|\Psi_{3,4}^{\pm}\rangle$ corresponding to $\sigma_{j}=\sigma_{j+1}=\mp1/2$
and $|\Psi_{3,4}^{0}\rangle$ corresponding to $\sigma_{j}=-\sigma_{j+1}$
which differ from each other only by the form of coefficient $b_{\pm}$.

In Appendix \ref{sec:Limiting-case} one can find the Ising limit
of the Eqs. \eqref{psi12} and \eqref{psi34}.

One of the unusual features of the eigenstates (\ref{psi12}) and
(\ref{psi34}) is the explicit dependance of the corresponding magnetic
moment on the magnetic field, which is a direct consequence of the
non-commutativity of the magnetization operator and block Hamiltonian.
It is easy to obtain that,
\begin{alignat}{1}
\mathcal{M}_{1,2}^{z}= & \langle\Psi_{1,2}|\left(g_{1}S_{j,1}^{z}+g_{2}S_{j,2}\right)|\Psi_{1,2}\rangle\label{M12}\\
= & \mp\frac{B(g_{1}-g_{2})^{2}}{4G},\nonumber
\end{alignat}
and
\begin{alignat}{1}
\mathcal{M}_{3,4}^{z}= & \langle\Psi_{3,4}|\left(g_{1}S_{j,1}^{z}+g_{2}S_{j,2}\right)|\Psi_{3,4}\rangle\label{M34}\\
= & \mp\frac{(g_{1}+g_{2})\left[B(g_{1}+g_{2})-2K(\sigma_{j}+\sigma_{j+1})\right]}{4F_{\sigma_{j},\sigma_{j+1}}}.\nonumber
\end{alignat}

Thus, $\mathcal{M}_{3,4}^{z}$ not only continuously depends on the
magnetic field but also exhibit jumps under the flip of the neighboring
$\sigma$-spins.

\subsection{Eigenstates for the chain}

Let us now describe the ground states for the whole chain using the
Hamiltonian \eqref{Hamdc1}, which are constructed with the aid of
the block eigenstates. In virtue of the difference in g-factors for
the Ising spins the model has six spins (two blocks) in the unit cell
and therefore the ground states will demonstrate the two-block translational
symmetry. Notice that in case of $g_{3}=g_{4}$, the unit cell can
contain only three spins (no period doubling).

\paragraph{1.- Quasi-Saturated (SQ) state:}

First of all, let us mention the quasi-saturated state with the corresponding
magnetic moment and energy per unit cell. Let us remind that the total number of the diamond-shaped blocks in the chain we denote by $N$, but due to the difference in the g-factors of the $\sigma$ spins the unit cell corresponding to the Hamiltonian \eqref{Hamdc1} contains six sites. Thus, all quantities presented below are calculated with respect to the number of the six-spin unit cells equal to $N/2$.

The first `quasi-saturated' ($QS_{1}$) state reads
\begin{alignat}{1}
|QS_{1}\rangle= & \prod_{j=1}^{\frac{N}{2}}|\uparrow\rangle_{_{2j-1}}\otimes|\Psi_{4}^{-}\rangle_{_{2j-1}}\otimes|\uparrow\rangle_{_{2j}}\otimes|\Psi_{4}^{-}\rangle_{_{2j}},\label{QS}\\
\mathcal{M}_{QS_{1}}= & \frac{(g_{1}+g_{2})\left(B(g_{1}+g_{2})-2K\right)}{2F_{+,+}}+\frac{1}{2}\left(g_{3}+g_{4}\right),\nonumber \\
E_{QS_{1}}= & \frac{\Delta}{2}-2F_{+,+}-\frac{B}{2}\left(g_{3}+g_{4}\right),\nonumber
\end{alignat}
here the arrow stand for the spin-up configuration of the corresponding
$\sigma$-spins.

The second `quasi-saturated' ($SQ_{2}$) state is expressed as follows:
\begin{alignat}{1}
|QS_{2}\rangle= & \prod_{j=1}^{\frac{N}{2}}|\downarrow\rangle_{_{2j-1}}\otimes|\Psi_{4}^{+}\rangle_{_{2j-1}}\otimes|\downarrow\rangle_{_{2j}}\otimes|\Psi_{4}^{+}\rangle_{_{2j}},\label{AF2}\\
\mathcal{M}_{QS_{2}}= & \frac{(g_{1}+g_{2})\left(B(g_{1}+g_{2})+2K\right)}{2F_{-,-}}-\frac{1}{2}\left(g_{3}+g_{4}\right),\nonumber \\
E_{QS_{2}}= & \frac{\Delta}{2}-2F_{-,-}+\frac{B}{2}\left(g_{3}+g_{4}\right).\nonumber
\end{alignat}

The eigenstates $QS_{1}$ and $QS_{2}$ are linked to each other by the inversion of the all $\sigma_j$ spins. The $QS_{2}$ represents a ground states at the strong magnetic field when the g-factors of the $\sigma$ spins are negative. Both $QS_{1}$ and $QS_{2}$ become degenerate at the vanishing magnetic field. They are the counterparts of the saturated or fully polarized state.
However, the $XY$-anisotropy $\gamma$ prevents the magnetization from
reaching its saturated value given by Eq.(\ref{mag0_dc}), at any
finite values of the magnetic field. Therefore, the saturation can
be reached asymptotically when $B\rightarrow\infty$ or at the vanishing
$XY$-anisotropy $\gamma\rightarrow 0$.
\paragraph{2.- Ferrimagnetic (FI) state: }

There are two 'ferrimagnetic' (FI) (with respect to the spin orientation, but not to the magnetic moment) eigenstates. This implies appearance of several sublattices with non-zero net magnetization as well as the nonzero $S^z$.

Thus the first ferrimagnetic ($FI_{1}$) state is
\begin{alignat}{1}
|FI_{1}\rangle= & \prod_{j=1}^{\frac{N}{2}}|\uparrow\rangle_{_{2j-1}}\otimes|\Psi_{2}\rangle_{_{2j-1}}\otimes|\uparrow\rangle_{_{2j}}\otimes|\Psi_{2}\rangle_{_{2j}},\label{F1}\\
\mathcal{M}_{FI_{1}}= & \frac{B(g_{1}-g_{2})^{2}}{2G}+\frac{1}{2}\left(g_{3}+g_{4}\right),\nonumber \\
E_{FI_{1}}= & -\frac{\Delta}{2}-2G-\frac{B}{2}\left(g_{3}+g_{4}\right).\nonumber
\end{alignat}

The unit cell of the ground state $FI_{1}$, thus, contains two Ising spin (with different g-factors) pointing up and
 two Heisenberg dimers with average two spin pointing up and two spin pointing down. Despite of non-coherent superposition of $|\uparrow\downarrow\rangle$ and $|\downarrow\uparrow\rangle$ in $|\Psi_{2}\rangle$ the expectation values of $S_{j,1}^1$ and $S_{j,2}^z$, though differ from $\pm 1/2$, compensate each other:
 \begin{eqnarray}
 &&_{_{j}}\langle\Psi_{2}|S_{j,1}^z|\Psi_{2}\rangle_{_{j}}=\frac 12\frac{1-c_-^2}{1+c_-^2}, \\
 &&_{_{j}}\langle\Psi_{2}|S_{j,2}^z|\Psi_{2}\rangle_{_{j}}=-\frac 12\frac{1-c_-^2}{1+c_-^2}. \\
 \end{eqnarray}
 However,
 \begin{eqnarray}
 _{_{j}}\langle\Psi_{2}|g_1 S_{j,1}^z|\Psi_{2}\rangle_{_{j}} \neq-\;_{_{j}}\langle\Psi_{2}|g_2 S_{j,2}^z|\Psi_{2}\rangle_{_{j}}.
 \end{eqnarray}
The second ferrimagnetic ($FI_{2}$) state is

\begin{alignat}{1}
|FI_{2}\rangle= & \prod_{j=1}^{\frac{N}{2}}|\downarrow\rangle_{_{2j-1}}\otimes|\Psi_{2}\rangle_{_{2j-1}}\otimes|\downarrow\rangle_{_{2j}}\otimes|\Psi_{2}\rangle_{_{2j}},\label{F3-1}\\
\mathcal{M}_{FI_{2}}= & \frac{B(g_{1}-g_{2})^{2}}{2G}-\frac{1}{2}\left(g_{3}+g_{4}\right),\nonumber \\
E_{FI_{2}}= & -\frac{\Delta}{2}-2G+\frac{B}{2}\left(g_{3}+g_{4}\right).\nonumber
\end{alignat}
Similarly, state $FI_{2}$ has two Ising spin-down,
two Heisenberg spin-down and two Heisenberg spin-up in the unit cell.

Despite the net spin orientation is not balanced, the magnetization
of the system could vanish at $g_{1}=g_{2}$ and $g_{3}=-g_{4}$.

Therefore, if one do not take into account the difference of
the g-factors ($g_{3}=g_{4}$) of the $\sigma$ spins these ground states
have three spins in the unit cell, "up-up-down" for the $FI_{1}$ state
and "down-up-down" for the $FI_{2}$ state. The pair of Heisenberg spins in both case form a perfect singlet state.

\paragraph{3.- Antiferromagnetic (AF) state:}

There are two eigenstates which one can call `antiferromagnetic' (AF), because the corresponding unit cell contains equal amount of spins pointing up and pointing down (balanced spin orientation). However, the magnetization does not necessarily
vanish, unless the particular case $g_{1}=g_{2}$ and $g_{3}=g_{4}$.

The first `antiferromagnetic' ($AF_{1}$) states is given by

\begin{alignat}{1}
|AF_{1}\rangle= & \prod_{j=1}^{\frac{N}{2}}|\uparrow\rangle_{_{2j-1}}\otimes|\Psi_{2}\rangle_{_{2j-1}}\otimes|\downarrow\rangle_{_{2j}}\otimes|\Psi_{2}\rangle_{_{2j}},\label{AF1}\\
\mathcal{M}_{AF_{1}}= & \frac{B(g_{1}-g_{2})^{2}}{2G}+\frac{1}{2}\left(g_{3}-g_{4}\right),\nonumber \\
E_{AF_{1}}= & -\frac{\Delta}{2}-2G-\frac{B}{2}\left(g_{3}-g_{4}\right).\nonumber
\end{alignat}

The second one ($AF_{2}$)  is
\begin{alignat}{1}
|AF_{2}\rangle= & \prod_{j=1}^{\frac{N}{2}}|\downarrow\rangle_{_{2j-1}}\otimes|\Psi_{2}\rangle_{_{2j-1}}\otimes|\uparrow\rangle_{_{2j}}\otimes|\Psi_{2}\rangle_{_{2j}},\label{AF1-1}\\
\mathcal{M}_{AF_{2}}= & \frac{B(g_{1}-g_{2})^{2}}{2G}-\frac{1}{2}\left(g_{3}-g_{4}\right),\nonumber \\
E_{AF_{2}}= & -\frac{\Delta}{2}-2G+\frac{B}{2}\left(g_{3}-g_{4}\right).\nonumber
\end{alignat}

The eigenstates $|AF_{1}\rangle$ and $|AF_{2}\rangle$ are slightly different
due to the left-right asymmetry which takes the place because of difference in g-factors for the Ising spins. They become identical when
$g_{3}=g_{4}$.

\paragraph{4.- Quantum ferrimagnetic (QI) state:}

Finally, we introduce two so--called `quantum ferrimagnetic` eigenstates. Here the number of the spin pointing up (down) in the unit cell is not a good quantum number (is not fixed), as the $|\Psi_{4}^{0}\rangle$ eigenstate for the quantum dimer is a non-coherent superposition of $|\uparrow\uparrow\rangle$ and $|\downarrow\downarrow\rangle$. Thus, the corresponding unit cell is characterized by one Ising spin pointing up, another one pointing down, and the total $S^z=0$ for the spins on Heisenberg dimers. But, in contrast to the AF and FI eigenstates it does not makes any sense to speak about the number of the Heisenberg spins with certain orientation even in the expectation value level.

The first `quantum ferrimagnetic` ($QI_{1}$) state reads
\begin{alignat}{1}
|QI_{1}\rangle= & \prod_{j=1}^{N/2}|\uparrow\rangle_{_{2j-1}}\otimes|\Psi_{4}^{0}\rangle_{_{2j-1}}\otimes|\downarrow\rangle_{_{2j}}\otimes|\Psi_{4}^{0}\rangle_{_{2j}},\label{QI1}\\
\mathcal{M}_{QI_{1}}= & \frac{B(g_{1}+g_{2})^{2}}{2F_{+,-}}+\frac{1}{2}\left(g_{3}-g_{4}\right),\nonumber \\
E_{QI_{1}}= & \frac{\Delta}{2}-2F_{+,-}-\frac{B}{2}\left(g_{3}-g_{4}\right),\nonumber
\end{alignat}
and the second `quantum ferrimagnetic` ($QI_{2}$) state differs from the previous one just by the orientation of the $\sigma$ spins,
\begin{alignat}{1}
|QI_{2}\rangle= & \prod_{j=1}^{\frac{N}{2}}|\downarrow\rangle_{_{2j-1}}\otimes|\Psi_{4}^{0}\rangle_{_{2j-1}}\otimes|\uparrow\rangle_{_{2j}}\otimes|\Psi_{4}^{0}\rangle_{_{2j}},\label{QI2}\\
\mathcal{M}_{QI_{2}}= & \frac{B(g_{1}+g_{2})^{2}}{2F_{+,-}}-\frac{1}{2}\left(g_{3}-g_{4}\right),\nonumber \\
E_{QI_{2}}= & \frac{\Delta}{2}-2F_{+,-}+\frac{B}{2}\left(g_{3}-g_{4}\right).\nonumber
\end{alignat}

Like in the previous case, the states $|QI_{1}\rangle$ and $|QI_{2}\rangle$
differ from each other only due to difference in the Ising spins g-factors.  They become identical when $g_{3}=g_{4}$. Note that if $g_{1}=-g_{2}$
and $g_{3}=g_{4}$ the magnetization can be zero.

In Appendix \ref{sec:Ground-state-I}  the ground states
energies for the limiting case of all Ising spins diamond chain are described.

\section{`Fire-Ice' interface}

In the Ref. \cite{yin15} an interesting unusual critical point has
been described. For the simplest classical case of the one-dimensional
ferromagnetic Ising model with staggered g-factors with different
signs the authors described the situation when there are two sub-lattice
(at zero temperature) in the ground state of the system. The one of
them is ordered and another one is totally disordered. For the obvious
reason they called the ground state `Half Fire, Half Ice'. However,
it is worthy mentioning, that the critical lines of aforementioned
kind have been considered a bit earlier. They are quite common properties
of the Ising-Heisenberg spin systems  \cite{tor16, bel14, str05, str03, can06, val08, can09, ant09, bel10, roj11a, oha12, bel13, gal13, ver13, gal14, ana14, tor14, qi14, lis15, abg15, gal15, gao15, lis16, hov16, ana17, rod17}.
Generally speaking, the ground states with ordered and disorder sublattices
naturally arise in the spin systems with complex unit cell containing
several spins. For instance, the same phenomena occurs in the ferromagnetic-ferromagnetic-antiferromagnetic
Ising chain, due to antiferromagnetic bond \cite{FFA03}. The appearance
of the antiferromagnetic bonds here is crucial for such critical states.
They usually arises as the degeneracy between two different ground
states which differ one from another by the orientation of one or
several spins. The simplest example can be found probably in Ref. \cite{can06}
in the Ising-Heisenberg S=1/2 diamond-chain. The critical line between
fully polarized state and the ground state where spins from quantum
dimer are pointing along the magnetic field, while the Ising spins
between them are pointing oppositely, due to antiferromagnetic coupling
between them and the dimers is the line corresponding to the 'one
third fire-two third ice' configuration. This means one sublattice
from three in the ground state is disordered. The principal difference
of the `Half Fire, Half Ice' configuration of the Ref. \cite{yin15}
from the partly ordered-partly disordered degenerate configurations
mentioned above is the uniform ferromagnetic coupling for all bonds.
The ambiguity in the state of the spins from the disordered sublattice
is here the consequence of their negative g-factor. As the model of
the Ising-Heisenberg diamond chain is the simplest generalization
of the Ising chain (decorated Ising chain) the corresponding 'fire-and-ice'
degenerate configurations also can be realized here.

Let as describe first how this configuration arises in the ordinary
ferromagnetic Ising chain with two alternating g-factors, $g_{A}>0$
and $g_{B}<0$ \cite{yin15}. The Hamiltonian is
\begin{equation}
\mathcal{H}_{Is}^{1d}=J\sum_{j=1}^{N}\sigma_{j}\sigma_{j+1}-B\sum_{j=1}^{N/2}\left(g_{A}\sigma_{2j-1}+g_{B}\sigma_{2j}\right),\label{Is_Ham1}
\end{equation}
where $J<0$, $\sigma_{j}=\pm1/2$ and we assume for simplicity, $|g_{B}|>g_{A}$.
Thus, for $T=0$ and sufficiently low magnetic field the ground state
will be ferromagnetic one with all spins pointed down. Then, it is
easy to see that there is a critical point at
\begin{eqnarray}
B_{c}=\frac{|J|}{g_{A}},\label{BcIs}
\end{eqnarray}
when the degeneracy occurs between aforementioned ground state and
a configuration where each spin with g-factor $g_{A}$ is pointed
up which becomes a non-degenerate ground state when $B>B_{c}$. Thus,
at the critical value of the magnetic field the system has two sub-lattices.
One of them is ordered (all spins with negative g-factor are pointed
down) and another one is completely disordered. This extraordinary
feature the authors of the Ref. \cite{yin15} named `Half Fire, Half
Ice'. This property can be easily obtained in the nearest-neighbor
ferromagnetic Ising model on arbitrary bipartite lattice with the
g-factors of different signs on each sublattice. Making the same assumptions
about $J$, $g_{A}$ and $g_{B}$, and considering the model with
two sublattices $A$ and $B$,
\begin{equation}
\mathcal{H}_{Is}^{AB}=J\sum_{i\in A,j\in B}\sigma_{i}\sigma_{j}-B\left( g_{A}\sum_{i\in A}\sigma_{i}+g_{B}\sum_{j\in B}\sigma_{j}\right),\label{Is_Ham2}
\end{equation}
one can easily see that there exists the zero-temperature critical
point of the same origin with the corresponding value of the magnetic
field
\begin{eqnarray}
B_{c}=\frac{|J|d}{2g_{A}},\label{Bcd}
\end{eqnarray}
where $d$ is the coordination number of the bipartite lattice.

It is easy first to do so for the purely Ising model on a diamond-chain,
then we can use the obtained results as a guideline for searching
the corresponding phenomena in the quantum model under consideration.

\subsection{Ising model on a diamond chain}

Let us consider the Ising limit of the Hamiltonian \eqref{Hamdc1}.
Assuming $J=0$ we get
\begin{alignat}{1}
\mathcal{H}_{I}= & \sum_{j=1}^{N}\left\{ \Delta S_{j,1}^{z}S_{j,2}^{z}+K(S_{j,1}^{z}\!+\!S_{j,2}^{z})(\sigma_{j}\!+\!\sigma_{j+1})\right.\nonumber \\
 & \left.-B(g_{1}S_{j,1}^{z}\!+\!g_{2}S_{j,2}^{z})-Bg_{j}\sigma_{j}\right\} ,\label{eq:Ising-H}
\end{alignat}
where the g-factors of the Ising intermediate spins alternating with
the spin-dimer are given by the Eq. \eqref{g34}.

The Ising diamond chain exhibits several interfaces between the ground states with peculiar partial
frustration. Below we discuss some of them using the results given
in appendix \ref{interface-state}. As the unit cell of the model contains six spins the "fire-and-ice" configurations with one, two, three, four and five frustrated (disordered) sublattice are possible. Let us emphasize once again that the origin of this frustration is the conflict between negative g-factor(s) and ferromagnetic couplings.

\subsubsection{One frustrated spin (1/6-fire and 5/6-ice)}

The interfaces with one frustrated spin in the unit cell are described in the Appendix \ref{interface-state}1. Here, we consider particular cases of negative $g_3$ as well as negative $g_3$ and $g_4$. All couplings are supposed to be ferromagnetic, $\Delta<0, K<0$. Let us first consider the case of only one spin with negative g-factor, say $g_3$. Due to ferromagnetic coupling between all spins the unit cell can be divided into two parts, the spin with negative g-factor, and the rest spins. When $|g_{3}|<2(g_{1}+g_{2})+g_{4}$ the zero-temperature ground state at the small enough magnetic filed is $QS_{1}^{+}$ (all spins are pointed up). The saturated state (the ground state at strong magnetic field) differs form $QS_{1}^{+}$ by the flip down of the spin with negative g-factor leading to $QI_{2}^{+}$ (See Eq. (\ref{B-QI2})). Thus, there is a critical value of the magnetic filed at which these two configurations become degenerate,
\begin{eqnarray}
B_{c}=\frac{2K}{g_{3}}.\label{BcIsDC4}
\end{eqnarray}
At this particular value of the magnetic field the system has one disordered and five ordered sites in the six-site unit cell. Using the terminology of the Ref. [\onlinecite{yin15}], one can naively refer to this state as to `1/6-fire-5/6-ice'. However, in virtue of the same argument about the entropy given above
the ground state belongs to the \textquotedbl{}half-fire-half-ice\textquotedbl{}
discussed in Ref.\cite{yin15}. The same situation can be also occurred at the interface between  $QS_{2}^{+}$ and $QI_{1}^{+}$ when both $g_3$ and $g_4$ are negative and additionally $|g_3|<|g_4|$. Let us emphasize that despite of triangular form
of the unit cell there is no geometrical frustration in the system as we have
only ferromagnetic bonds. The appearance of disorder here is
the direct consequence of the interplay between the ferromagnetic interaction and
negative g-factors \cite{yin15}.

\subsubsection{Two frustrated spins (1/3-fire and 2/3-ice)}

All possible interfaces with two frustrated spins are described in the Appendix \ref{interface-state}2. As an illustration here we analyze particular case, $g_3<0$ and $g_4<0$ for ferromagnetic coupling. Consider first the case $|g_3+g_4|<2(g_1+g_2)$. Thus, the interface with two disordered sublattices arises between $QS_1^+$ and $QS_2^+$ at
\begin{eqnarray}
B_{c}=\frac{4K}{g_{3}+g_{4}}.\label{BcIsDC2}
\end{eqnarray}
As here we have two spins from the six-spin unit cell disordered, according to our convention \cite{yin15} the corresponding state can be referred to as to `1/3-fire and 2/3-ice' configuration. At $|g_3+g_4|>2(g_1+g_2)$ the interface with two disordered sublattices occurs between $QS_2^-$ and $QS_2^+$ at the particular value of the magnetic field,
\begin{eqnarray}
B_{c}=\frac{2|K|}{g_{1}+g_{2}}.\label{BcIsDC2_2}
\end{eqnarray}

Let us now consider the case of antiferromagnetic coupling into the dimer, $\Delta>0$ and $K<0$.  Assuming for the time being $g_1=g_2>0$ and $g_3<0, g_4<0$
we can find the ground state of the Ising diamond-chain at weak magnetic
field pointing along $z$-axis to be four-fold degenerate (or two-fold degenerate when $g_3=g_4$ and the magnetic unit cell shrunk to the three-spin plaquette): $\sigma_{j}=-1/2,S_{j,1}^{z}=\pm1/2$,
$S_{j,2}^{z}=\mp1/2$. But the degeneracy is lifted once we put
$g_{1}\neq g_{2}$. The degeneracy can be even sixteen-fold at zero magnetic field when $\sigma$-spins become frustrated as well. The corresponding ground state is the Ising counterpart of the dimer-monomer ground state of the quantum diamond chain \cite{kikuchi03}. Supposing as well $\Delta>|K|$ we get the non-degenerate ground
state with four spins in the six-spin unit cell pointing down ($\sigma_{j}=\sigma_{j+1}=-1/2$
and $S_{j,2}^{z}=S_{j+1,2}^{z}=-1/2$) and two spins pointing up ($S_{j,1}^{z}=S_{j+1,1}^{z}=1/2$), or $FI_2^+$ (See Eq. (\ref{B7})). In the saturated state ($QS_{1}^{+}$) at strong magnetic field one has the flip-up of the spins with g-factor $g_2$. Therefore, there is a critical value
of the magnetic field, defining the interface $QS_{1}^{+}\leftrightarrow FI_{1}^{+}$
($QS_{2}^{+}\leftrightarrow FI_{2}^{+}$):
\begin{eqnarray}
B_{c}=\frac{\Delta+2|K|}{2g_{2}}.\label{BcDC2}
\end{eqnarray}
 Thus, we defined
here another `fire-and-ice' ground sates, in which four spins in the
unit cell are ordered and two is completely disordered. In virtue
of the distribution of the spins from the unit cell between ordered
and disordered sublattices, we can refer to this ground state as to
`1/3-fire-2/3-ice'.

\subsubsection{Three frustrated spins (1/2-fire and 1/2-ice)}

Increasing further the number of the frustrated sites in the unit cell one can arrive at the ground states
 given in appendix \ref{interface-state}3.

As an example we consider the following distribution of the g-factors for the ferromagnetic
Ising diamond-chain: $g_{1}$<0 and $g_{3}$<0, while $g_{2}>0$ and $g_{4}>0$.
 Two different regimes $|2g_{1}+g_{3}|>2g_{2}+g_{4}$
and $|2g_{1}+g_{3}|<2g_{2}+g_{4}$ leading to  different value of the critical
field are possible. However, in both situation there are three ordered and three \textquotedbl{}coherently\textquotedbl{} disordered spins in the six-spin
 magnetic unit cell. Therefore, we
deal here with exact \textquotedbl{}half-fire-half-ice\textquotedbl{} configuration
with the values of the critical field in the interface between $QS_2^-$ and $AF_2^+$ for $|2g_1+g_3|>2g_2+g_4$ and
$QS_1^+$ and $AF_2^-$ for $|2g_1+g_3|>2g_2+g_4$. The corresponding values of the magnetic field are
\begin{eqnarray}
B_{c}=\frac{|\Delta|+2|K|}{2g_{2}+g_{4}}.\label{BcIsDC5}
\end{eqnarray}
and
\begin{eqnarray}
B_{c}=\frac{\Delta+2K}{2g_{1}+g_{3}},\label{BcIsDC6}
\end{eqnarray}
respectively. It is worth emphasizing once again that the three disordered spins in the unit cell can not be disordered independently of each other, they can change their direction only simultaneously. We call them, thus, \textquotedbl{}coherently\textquotedbl{} disordered spins. Therefore, the residual entropy per the unit cell at the critical values of the magnetic field is equal to $\log 2$.

\subsubsection{Four frustrated spins (2/3-fire and 1/3-ice)}

All possible interfaces between two ground eigenstates with four frustrated spins in the unit cell are presented in the Appendix \ref{interface-state}4.

As an example we consider the following case:
$\Delta<0$, $K<0$ and $g_{3}<0$, $g_{4}<0$, $g_{1}>0$,
$g_{2}>0$. Again the ground state at sufficiently weak magnetic field depends on the mutual relation of total positive and total negative g-factors in the unit cell. For $|g_{3}+g_{4}|>2(g_{1}+g_{2})$ the system demonstrate all spins pointing down, or $QS_2^-$.
Then, for strong enough magnetic field the ground state transforms into the saturated one, which in this case is $QS_2^+$
The degeneracy between these ground states takes place at
\begin{eqnarray}
B_{c}=\frac{2|K|}{g_{1}+g_{2}},\label{BcIsDC1}
\end{eqnarray}
As there are four \textquotedbl{}coherently\textquotedbl{} disordered spins and two ordered spins in the magnetic unit cell, the corresponding configuration can be classified as "2/3-fire and 1/3-ice".
If the g-factors of $\sigma_{j}$-spin equal to each
other, $g_{3}=g_{4}$,  the magnetic unit cell shrunk to only three spins. It is also straightforward that
this critical value is nothing else but the Eq. (\ref{Bcuplo}) with
$K<0$ and $\sigma_{j}=\sigma_{j+1}=-1/2$.

\subsubsection{Five frustrated spins (5/6-fire and 1/6-ice)}

Finally, the last case we present here is the interface with only one order site from the six sites in the magnetic unit cell. One
can consider the following case $g_{3}<0$ with all other g-factors
positive. Again, the ratio of ordered and disordered sites inside
the magnetic unit cell depends on the relation between $|g_{3}|$
and $2(g_{1}+g_{2})+g_{4}$.

If $|g_{3}|>2(g_{1}+g_{2})+g_{4}$ then the low-field orientation of
all six spins form the magnetic unit cell is defined by the orientation
of spin with negative g-factor. In the saturated state all spins except
the one with negative g-factor are pointed up. Thus, at the critical
value of the magnetic field arises the boundary  $QS_{1}^{+}\leftrightarrow QI_{1}^{-}$
and $QS_{1}^{-}\leftrightarrow QI_{1}^{+}$
\begin{eqnarray}
{\normalcolor B_{c}={\normalcolor \frac{2|K|}{g_{4}-2(g_{1}+g_{2})},}}\label{BcIsDC3}
\end{eqnarray}
the system has only one ordered site among the six sites of the magnetic
unit cell. Thus, the corresponding configuration could be referred
to as `5/6 fire-1/6 ice'. However, it is important to remember that
the rest five spins from the unit cell can not be disordered independently
form each other. For each magnetic unit cell they all can either point
up or down. Therefore, the residual entropy per the unit cell is $\log 2$.

\subsubsection{Additional remarks}
Another important point affecting the structure of the partially frustrated interface is the relation between the total negative and positive g-factors in the system. Consider arbitrary S=1/2 Ising spin lattice with the uniform ferromagnetic coupling $K$. It can be naively seemed that the number of the frustrated spins into the magnetic unit cell in the interface between two ground states corresponds to the number of the spins with negative g-factor in it. However, this is true only when absolute value of the total negative g-factor is bigger than the total positive g-factor for the unit cell, $\sum|g_{neg}|>\sum g_{pos}$. The ground state under this condition (at infinitely low magnetic field in $z$ direction) is the ferromagnetic state with all spins pointing down, $F_-$. At the opposite limit, $\sum|g_{neg}|<\sum g_{pos}$, the ground state is ferromagnetic with all spins pointing up, $F_+$. The saturated state corresponds to all spins with positive g-factor pointing up and all spins with negative g-factor pointing down, $S$. It is easy to write down the corresponding ground states energies per one unit cell in the following form:
\begin{eqnarray}
&& E_F^-=-\frac{n}{4} K+\frac{B}{2}\left(\sum g_{pos}-\sum|g_{neg}|\right),\\
&& E_F^+=-\frac{n}{4} K-\frac{B}{2}\left(\sum g_{pos}-\sum|g_{neg}|\right),\nonumber\\
&&E_S=-\frac{m}{4} K-\frac{B}{2}\left(\sum g_{pos}+\sum|g_{neg}|\right),\nonumber
\end{eqnarray}
 where $n$ and $m$ are positive integers and $m<n$. When $\sum|g_{neg}|>\sum g_{pos}$ the interface between $F_-$ and $S$ exists at
 \begin{eqnarray}\label{Bcpos}
 B_c=\frac{(n-m)K}{4\sum g_{pos}}.
 \end{eqnarray}
 Thus, all spins with positive g-factor are "coherently" frustrated and the critical field does not depend on the rest of g-factors. For the opposite situation, $\sum|g_{neg}|<\sum g_{pos}$, the ordered and disordered sites change over each other. Now, at the interface between $F_+$ and $S$ at the critical value of the magnetic field,
 \begin{eqnarray}\label{Bcneg}
 B_c=\frac{(m-n)K}{4\sum|g_{neg}|},
 \end{eqnarray}
 all spins with positive g-factors are ordered within the magnetic unit cell, while the spins with negative g-factors are "coherently" disordered. Thus, one can conclude that having $q$ spins with negative g-factors in the p-spin magnetic unit cell can lead to the interface with either $q$ disordered spins or $p-q$ disordered spins in the unit cell depending on the relation between total negative and positive g-factors. Let us illustrate this feature on the example of one and five disordered sites in the unit cell for the Ising diamond chain. Eqs. (\ref{BcIsDC4}) and (\ref{BcIsDC3}) are the examples of Eqs. (\ref{Bcneg}) and (\ref{Bcpos}) respectively. Taking, $g_3>0$ and the rest g-factors negative one can obtain the same interfaces given by the Eqs. (\ref{BcIsDC4}) and (\ref{BcIsDC3}) but under the opposite relation between the g-factors. Thus, when $g_3>|2 (g_2+g_2)+g_4|$ one gets "5/6-fire-1/6-ice" with five disorder spins in the unit cell and the critical field given by the Eq. (\ref{BcIsDC3}). For $g_3<|2 (g_2+g_2)+g_4|$ the system has the interface corresponding to "1/6-fire-5/6-ice" with one disordered spin and critical value of magnetic field coinciding with Eq. (\ref{BcIsDC4}).

\subsubsection{The phase diagrams}

Let us now proceed to the description of the ground state phase diagrams illustrating
the interfaces discussed above as well as more sophisticated cases connected with interplay between negative g-factors and antiferromagnetic and mixed coupling in the system. In the examples presented above we considered the simplest interfaces between ground states at low (vanishing) magnetic field and the saturated state at strong enough magnetic filed. Here, using the phase diagrams we demonstrate some intermediate interfaces as well.

\begin{figure}
\includegraphics[scale=0.3]{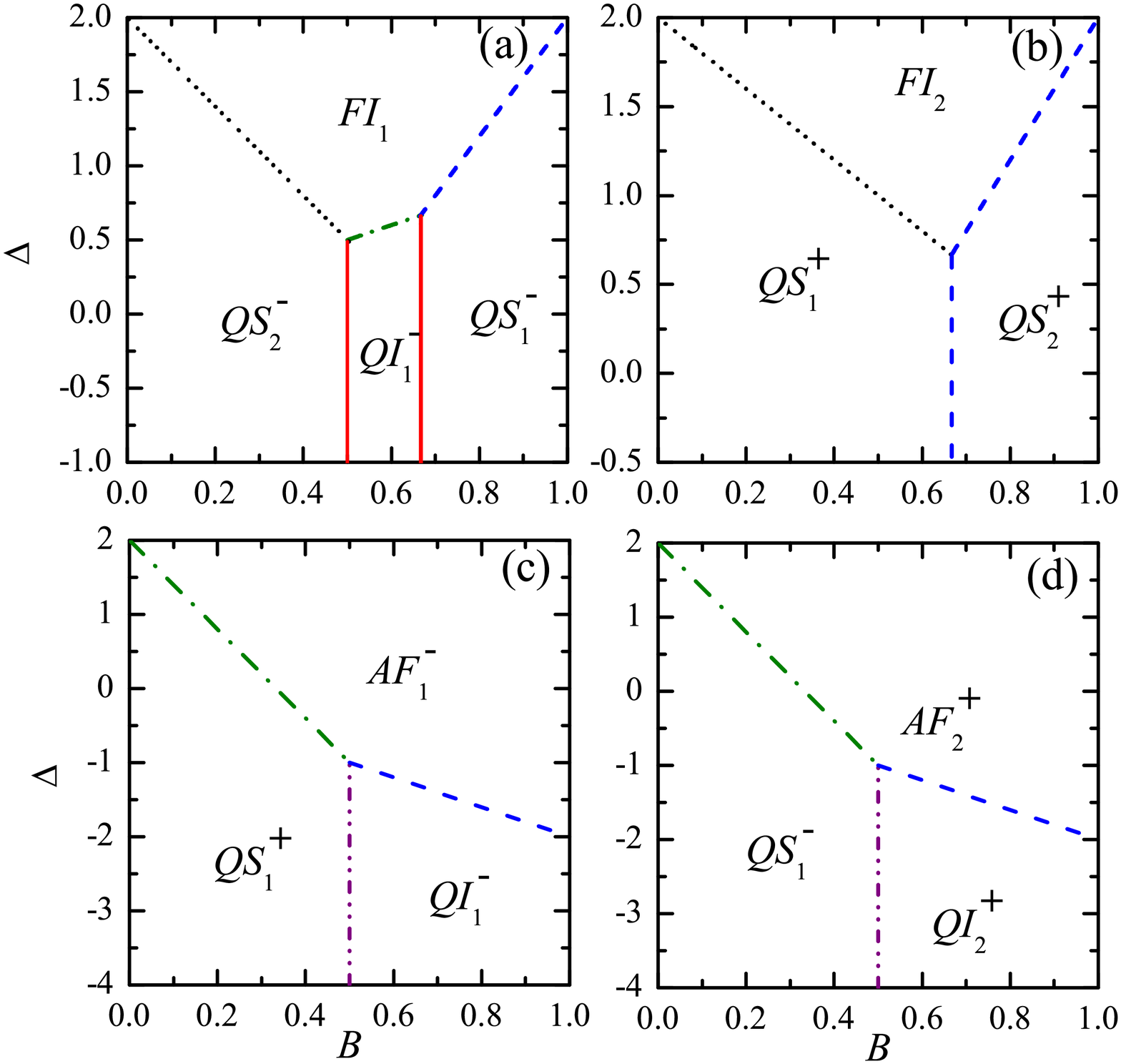}\caption{\label{phd-Ising}Zero temperature phase diagram $\Delta$ against
$B$. (a) For $K=-1$, $g_{1}=-2$, $g_{2}=-2$, $g_{3}=4$ and $g_{4}=3$.
(b) For $K=-1$, $g_{1}=2$, $g_{2}=2$, $g_{3}=-3$ and $g_{4}=-3$.
(c) For $K=-1$, $g_{1}=-2$, $g_{2}=1$, $g_{3}=4$ and $g_{4}=-2$.
(d) $K=1$, $g_{1}=2$, $g_{2}=-1$, $g_{3}=-2$ and $g_{4}=4$.}
\end{figure}

The zero-temperature ground states phase diagrams in the plane
$B$-$\Delta$ for the Ising diamond-chain with the Hamiltonian \eqref{eq:Ising-H} are
presented in the Fig.\ref{phd-Ising}. The panel (a)  displays the phase diagram corresponding to the following particular values of the parameters: $K=-1$,
$g_{1}=-2$, $g_{2}=-2$, $g_{3}=4$ and $g_{4}=3$. Here one can see five
interfaces between four ground states, $QS_2^-$, $QI_1^-$, $QS_2^-$ and $FI_1$, the latter is degenerate superposition of $FI_1^+$  and $FI_1^-$, which exists due to the equal g-factors of the spins from vertical dimer. Moreover, the spins in each vertical dimer can be either in "up-down" or in "down-up" configuration independently even inside the unit cell. Thus, the corresponding state has the residual entropy per the unit cell, $\mathcal{S}=2 \log 2$.

The interface $QS_{2}^{-}\leftrightarrow FI_{1}$ corresponds
to `2/3 fire-1/3 ice' state, with residual entropy per the unit cell, $\mathcal{S}= 2\log 2$,
although there are four frustrated spins these spins are not frustrated independently from each other. The interface $QS_{1}^{-}\leftrightarrow FI_{1}$ corresponds to `1/3 fire-2/3 ice' state. It has residual entropy per the unit cell, $\mathcal{S}=2\log 3$. The degeneracy is so high due to independent possibility for for two vertical dimers form the six-spin unit cell to be in one of the three spin configurations, "up-down", "down-up" and "down-down" at the $QS_{1}^{-}\leftrightarrow FI_{1}$  interface.
Another interface is $QI_{1}^{-}\leftrightarrow FI_{1}$. It corresponds to `1/2 fire-1/2
ice' state with residual entropy the unit cell, $S=\log 5$. Quite remarkable is a series of two interfaces, $QS_{2}^{-}\leftrightarrow QI_{1}^{-}$ and $QS_{1}^{-}\leftrightarrow QI_{1}^{-}$, which both correspond to the `1/6 fire-5/6 ice' configuration, but with the different frustrated sites. The residual entropy per the unit cell in both cases is $\mathcal{S}=\log 2$. Thus, one can see that the intermediate eigenstate $QI_{1}^{-}$ between the ground state at zero magnetic field and the saturated state arises due to difference in the g-factors of two spins between the dimers.

In the Fig.\ref{phd-Ising}(b),
we present the phase diagram for the case $g_1=g_2>0$, $g_3=g_4<0$ for ferromagnetic coupling $K=-1$. Under these conditions for the g-factors we have only three sites in a unit cell.

Here one can see three phases and three interfaces. The interface between  $QS_{1}^{+}$ and $QS_{2}^{+}$ represents
 `1/3 fire-2/3 ice' configuration with residual entropy per the unit cell $\mathcal{S}=\log 2$ \footnote{Let us remind the reader that in this particular choice of the g-factors the system has only three spins in the magnetic unit cell. That is why we get $\log 2=\frac{\log 4^{\frac N2}}{N}$ instead of $2\log2$ for the case of six-spin unit cell.},
similarly there is the boundary between $FI_{2}$ and $QS_{2}^{+}$
states corresponding to `2/3 fire-1/3 ice' state with residual entropy the unit cell
$\mathcal{S}=\log 3$. The interface between $QS_{1}^{+}$ and $ FI_{2}$ is characterized by
 four frustrated spins generating the so-called `2/3 fire-1/3
ice' state with residual entropy per the unit cell,$\mathcal{S}=\log2$.

    In the Fig.\ref{phd-Ising}(c) for $g_1=-2, g_2=1, g_3=4$ and $g_4=-2$ one can see additional type of interface
$QS_{1}^{+}\leftrightarrow QI_{1}^{-}$ corresponding to `5/6 fire-1/6 ice' state. However, as in this case the total magnetic moment of the unit cell is zero, $2(g_1+g_2)+g_3+g_4=0$ we have simple (non-macroscopic) degeneracy between $QS_{1}^{+}$ (all spins up) and $QS_{2}^{-}$ (all spins down). The  vanishing total g-factor within the unit cell also leads to a very high degeneracy at the horizontal line $B=1/2$. The local mixture of the $QS_{1}^{+}$, $QS_{2}^{-}$ and $QI_{1}^{-}$ states yields the asymptotic value of the residual entropy per unit cell in the thermodynamic limit, $\mathcal{S}=\log(\frac{3+\sqrt{5}}{2})$\cite{tor16}.
There is another interface between $QI_{1}^{-}$ and $ AF_{1}^{-}$ representing the
 `1/3 fire-2/3 ice' configuration with residual entropy given
by $\mathcal{S}=2\log2$.
The interfaces $QS_{1}^{+}\leftrightarrow AF_{1}^{-}$
corresponds to `1/2 fire-1/2 ice' configuration and has residual entropy per block $\mathcal{S}=\log(\frac{3+\sqrt{5}}{2})$.
Finally, the Fig.\ref{phd-Ising}(d)  is quite similar to the previous one.

\subsection{Ising-Heisenberg diamond chain}

Let us now turn to our Ising-Heisenberg model and look for the quantum
or semi-classical counterparts of the \textquotedbl{}fire-and-ice\textquotedbl{}
configurations described above for the purely Ising case.

Concerning the Ising-Heisenberg diamond-chain we considered here the
same phenomena takes place. We have to put $\gamma=0$ as any non-zero
XY-anisotropy mixed up the \textquotedbl{}up-up\textquotedbl{} and
\textquotedbl{}down-down\textquotedbl{} states for the quantum spin
dimer. Thus, the \textquotedbl{}disorder\textquotedbl{} term being
applied to the separate spins of the quantum dimer to some extent
makes no sense. However, the degeneracy still can exist. For exchange
parameters $J<0$, $K<0$ and $\Delta>0$, with g-factors $g_{1}=g_{2}>0$,
$g_{3}=g_{4}<0$, and $|g_{3}|>2g_{1}$ the degeneracy occurs between
$|QS_{2}\rangle$ and the eigenstate with all spins pointing down
at the value of the magnetic field given by Eq. (\ref{BcIsDC1}).
Moreover, at $\Delta=1$ the degeneracy rises up as the energy of
the eigenstate $|FI_{1}\rangle$ at the critical value of the magnetic
field $B_{c}$ become equal to those of $|QS_{2}\rangle$ and $|QS_{1}\rangle$.
However, this setup is just simple generalization
of the Ising chain from Ref. \cite{yin15}.

In case of the quantum dimer the situation is different. For $g_{1}=g_{2}$
instead of \textquotedbl{}up-up\textquotedbl{} and \textquotedbl{}down-down\textquotedbl{}
configurations we have the spin-singlet (for $g_{1}=g_{2}$) and the
$S_{tot}^{z}=0$ component of the triplet. that is why the full analog
of the classical `1/3 fire-2/3 ice` state in quantum case does not
exist. We are going to consider instead the analog of the critical
line between fully polarized eigenstate and the eigenstate with Ising
spins pointing down. As was mentioned above the corresponding state
caused by the antiferromagnetic coupling $K$ has been considered
in Ref. \cite{can06} . In our case, however, we consider ferromagnetic
$K$ and negative $g_{3}$ and $g_{4}$ with interface between $|QS_{1}\rangle$
and $|QS_{2}\rangle$. One has to keep in mind that the case of the
negative g-factors of the $\sigma$-spins corresponds to the saturated
state $|QS_{2}\rangle$, while $|QS_{1}\rangle$ has intermediate
magnetization. The critical $B_{c}$ can be find from eqs. \eqref{QS}
and \eqref{AF2}:
\begin{alignat}{1}
B(g_{3}+g_{4})= & \sqrt{\left(B(g_{1}+g_{2})+2K\right)^{2}+J^{2}\gamma^{2}}\nonumber \\
 & -\sqrt{\left(B(g_{1}+g_{2})-2K\right)^{2}+J^{2}\gamma^{2}}.
\end{alignat}
Solving this equation we get
\begin{equation}
B_{c}=\sqrt{\frac{16K^{2}}{(g_{3}+g_{4})^{2}}+\frac{4J^{2}\gamma^{2}}{(g_{3}+g_{4})^{2}-4(g_{1}+g_{2})^{2}}}.\label{eq:q-Bc}
\end{equation}

\begin{figure}[h]
\includegraphics[scale=0.3]{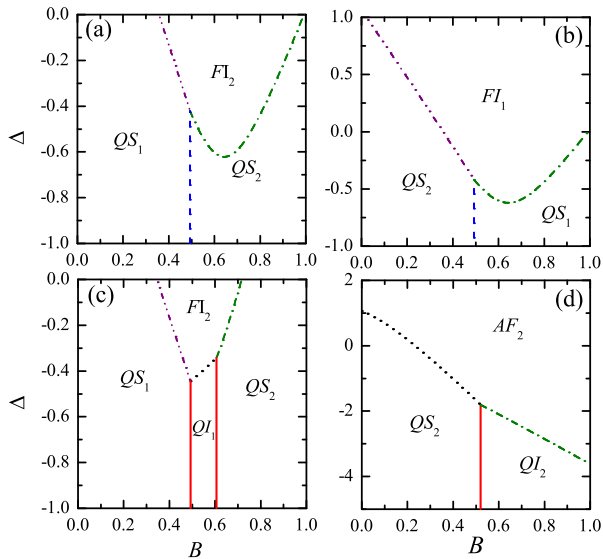}\caption{\label{qph-dgm1}Zero temperature phase diagrams for the Ising-Heisenberg diamond chain at fixed $J=1$ and $\gamma=0.5$. (a) For $K=-1$, $g_{1}=2$,
$g_{2}=1.2$ , $g_{3}=-3$ and $g_{4}=-3$. (b) For $K=1$, $g_{1}=2$,
$g_{2}=1.2$, $g_{3}=3$ and $g_{4}=3$ . (c) For $K=-1$, $g_{1}=2$,
$g_{2}=2$, $g_{3}=-3$ and $g_{4}=-4$. (d) For $K=-1$, $g_{1}=-2$,
$g_{2}=2$, $g_{3}=-4$ and $g_{4}=3$.}
\end{figure}

In the Fig.\ref{qph-dgm1} the ground states zero temperature phase diagrams in the
($B$-$\Delta$)-plane are presented for fixed $J=1$ and $\gamma=0.5$. The critical line given by the Eq. (\ref{eq:q-Bc}) can be found in three panels, (a) and (b). In the panels (c) and (d) one can see another interfaces given by the vertical lines, $QS_1\leftrightarrow QI_1$ and $QS_2\leftrightarrow QI_1$.   In the panel (a) the following values of teh parameters are chosen: $g_{1}=2$, $g_{2}=1.2$,
$K=-1$, $g_{3}=-3$ and $g_{4}=-3$. The interface
between $QS_{1}$ and $QS_{2}$ corresponding to `1/3 fire-2/3 ice'
state, with critical $B_{c}$ given by \eqref{eq:q-Bc} resulting
in $B_{c}=0.4927794$. At first glance this interface seems to be
frustrated, analogous to the Ising limit case (see Fig.\ref{phd-Ising}(b)),
but due to the fact that the states $|\Psi_{4}^{-}\rangle=0.9975898|\uparrow\uparrow\rangle-0.06938726|\downarrow\downarrow\rangle$
and $|\Psi_{4}^{+}\rangle=0.4207327|\uparrow\uparrow\rangle-0.9071847|\downarrow\downarrow\rangle$
are different, the Ising-Heisenberg diamond chain is simply two-fold
degenerate and not macroscopically degenerate, thus there is no residual
entropy in this interface. However, there exists a quantum frustrated
interface between $QS_{2}\leftrightarrow FI_{2}$ with residual entropy
$\mathcal{S}=2\log 2$ and another quantum frustrated interface
$QS_{1}\leftrightarrow FI_{2}$ with a non-trivial residual entropy
depending of the parameters $B$ and $\Delta$. In the Fig.\ref{qph-dgm1}b
 the phase diagram for fixed $K=1$, $g_{1}=2$, $g_{2}=1.2$,
$g_{3}=3$ and $g_{4}=3$ is presented. Here we observe once again the interface
$QS_{1}\leftrightarrow QS_{2}$ whose critical magnetic field is given
by \eqref{eq:q-Bc} ($B_{c}=0.4927794$). Similarly, there is also
the quantum frustrated interface between $QS_{2}\leftrightarrow FI_{1}$
with residual entropy $\mathcal{S}=2\log 2$ and another quantum
interface $QS_{1}$ and $FI_{1}$ with a non-trivial residual
entropy depending of the parameters $B$ and $\Delta$. In the Fig.\ref{qph-dgm1}c
one can see the phase diagram for fixed $J=1$, $K=-1$, $\gamma=0.5$, $g_{1}=2$ and $g_{2}=2$.
Here the following phases are presented: $QS_{1}$, $QI_{1}$ , $QS_{2}$
and $FI_{2}$. Two interfaces $QS_{1}\leftrightarrow QI_{1}$
and $QI_{1}\leftrightarrow QS_{2}$ corresponding to `1/6 fire-5/6 ice'
state have residual entropy given by $\mathcal{S}=\log2)$. Whereas,
the interface $QS_{2}\leftrightarrow FI_{2}$ is frustrated with residual
entropy given by $\mathcal{S}=2 \log2$. The other interfaces
are pure quantum frustrated states. Similarly in Fig.\ref{qph-dgm1}d
 the phase diagram for fixed $J=1$, $K=-1$, $\gamma=0.5$, $g_{1}=-2$,
$g_{2}=2$, $g_{3}=-4$ and $g_{4}=3$ is shown. The interface
between $QS_{2}$ and $QI_{2}$ representing the `1/6 fire-5/6 ice'
state has residual entropy given by $\mathcal{S}=2\log2$.
The other interfaces $QS_{2}\leftrightarrow AF_{2}$ and $QI_{2}\leftrightarrow AF_{2}$
are pure quantum frustrated states with residual entropies given by $\mathcal{S}=\log2$.
$\mathcal{S}=2\log2$ respectively.

\subsection{Magnetization and magnetic Susceptibility}

The fact that the magnetization is a non-conserved quantity gives
rise to an unusual behavior of the magnetic susceptibility at zero
temperature.

\begin{figure}
\includegraphics[scale=0.31]{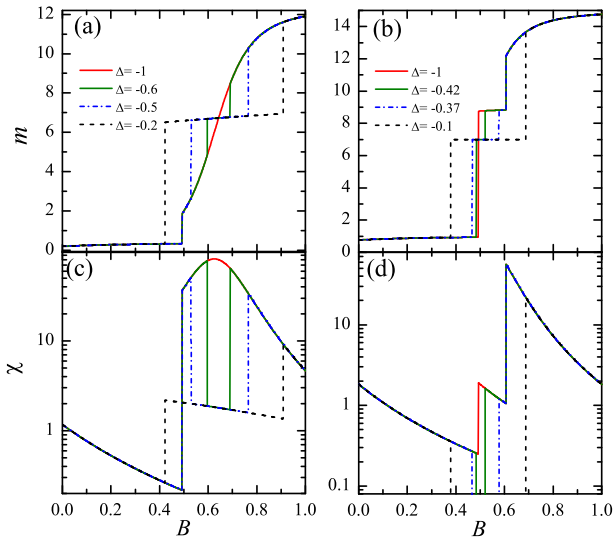}

\caption{\label{fig:Mag-X}Zero temperature magnetization (magnetic susceptibility)
as a function of magnetic field $B$ for $J=1$, $K=-1$ and
$\gamma=0.5$. Left panel, (a) and (c), $g_{1}=2$, $g_{2}=1.2$, $g_{3}=-3$ and
$g_{4}=-3$; roght panel, (b) and (d),  For $g_{1}=2$, $g_{2}=2$, $g_{3}=-3$ and $g_{4}=-4$.}
\end{figure}

In the Fig.\ref{fig:Mag-X} the magnetization and magnetic susceptibility
as a function of $B$ is displayed for fixed values of the parameters $J=1$, $K=-1$ and
$\gamma=0.5$. The left panel, Fig.\ref{fig:Mag-X}a and  Fig.\ref{fig:Mag-X}c show the magnetization and susceptibility
for $g_{1}=2$, $g_{2}=1.2$, $g_{3}=-3$ and $g_{4}=-3$. The effects of the
 the non-conserving magnetization are well visible here. For low magnetic field
there is a magnetization quasi-plateau \cite{oha15,bel14} which gives the impression
of a constant value of the magnetization. However,  actually this region corresponds to the state
$QS_{1}$ with non-conserving magnetization which has weak but monotone dependence of the magnetic field.
 The magnetization curves for $\Delta=-0.5$ and $\Delta=-0.6$  demonstrate another quasi-plateau at $B\approx0.641$ corresponding to $FI_{2}$. The final part of the curve corresponds to the quasi-saturates state, $QS_{2}$.
 In the panel (c)
the magnetic susceptibility for the same set of parameters is shown. The behavior of the susceptibility evidences of the non-plateau nature of the magnetization within the same eigenstates of the system. The interesting feature is monotone decrease of the susceptibility as a function of the magnetic field for the $QS_{1}$ eigenstate and non-monotone behavior with the maximum for the $QS_{2}$.

The right panel of the  Fig.\ref{fig:Mag-X} demonstrate the magnetization and susceptibility
as a functions of the magnetic field for $J=1$, $K=-1$ and $\gamma=0.5$
with g-factors $g_{1}=2$, $g_{2}=2$, $g_{3}=-3$ and $g_{4}=-4$.
Again here one can find the quasi-plateau corresponds to
the state $QS_{1}$ for low magnetic field and the quasi-saturated state $QS_{2}$ for the strong field.
 There is also an intermediate quasi-plateau, which corresponds to the $FI_{2}$ state for $\Delta=-0.1$. The third quasi-plateau arises for for $\Delta=-0.37$ and $\Delta=-0.42$. This is the $QI_{1}$ state which follows the $FI_{2}$ in these cases.
However,  for $\Delta=-1$ there is only one intermediate quasi-plateau (second one) corresponding to the $QI_{1}$ state. The corresponding magnetic susceptibility is shown in the panel (d). Here, it demonstrates the monotone decrease for each quasi-plateau.

\section{Conclusion}

In this paper we consider the Ising-XYZ model on the diamond chain
which was assembled as follows: the particles with the Ising
spin are located at the nodal diamond chain sites, whereas Heisenberg
spins are over interstitial sites. We have assumed the Ising spin
and the Heisenberg spins have different g-factors, as well as we have
assumed the system is under external magnetic field. The non-commutativity
of the magnetization operator and Hamiltonian is due to the different
g-factors of Ising and Heisenberg spins and due to the  XY-anisotropy ($\gamma$)
in the Heisenberg exchange interaction. This leads to the unusual
phenomena, such as the non-linear magnetic field dependence of the spectrum and non-constant magnetization within the same ground state.
We discuss in detail the zero temperature phase diagram under several
conditions and we find interesting phases. Due to the non-uniform sighs of the four g-factors presented in the unit cell
there are a phase boundaries corresponding to so-called ``half
fire-half ice\textquotedblright{} configurations for ferromagnetic couplings, which contain ordered and disordered sublattices simultaneously. These interfaces for Ising diamond were classified in five groups: the first one is
when there is one frustrated (disordered) spin and rest five ordered spins in the unit
cell ( `1/6 fire-5/6 ice'); similarly for two frustrated
spins ( `1/3 fire-2/6 ice'); for three frustrated spins
(`1/2 fire-1/2 ice'), for four frustrated spins ( `2/3 fire-1/3 ice'),
and finally for five frustrated spins ( `5/6 fire-1/6 ice').
For quantum Ising-Heisenberg diamond chain most of the interfaces
become quantum frustrated states.

Besides that, we study the zero temperature magnetization and magnetic
susceptibility as a function of the external magnetic field.

\section{Acknowledgements}

V. O. acknowledges the partial financial
support form the grants by the State Committee of Science of Armenia
No. 15-1F332 and SFU-02 as well as from the ICTP Network NT-04. He also expresses his deep gratitude to the ICMP (Lviv) for warm hospitality during the final stage of the work. O. R. thanks the Brazilian
agencies FAPEMIG and CNPq for partial financial support.

\appendix

\section{\label{sec:Limiting-case}Limiting case $\gamma=0$ and $g_{1}-g_{2}=0$}

Although, the case of conserving magnetization ($\gamma=0$ and $g_{1}-g_{2}=0$)
is trivial and well known, we are going to make few comments about
these limits and some new notations which are suitable for our further
analysis. First of all let us mention, the limit $g_{1}-g_{2}=0$
under which the $|\Psi_{1,2}\rangle$ transform into the conventional
singlet and zeroth component of triplet for two spin-1/2:
\begin{eqnarray}
 &  & |\Psi_{1,2}\rangle\rightarrow\frac{1}{\sqrt{2}}\left(|\uparrow\downarrow\rangle\pm\mbox{sgn}(J)|\downarrow\uparrow\rangle\right),\label{Psi12lim}\\
 &  & \varepsilon_{1,2}=-\frac{\Delta}{4}\pm\frac{1}{2}|J|.\nonumber
\end{eqnarray}
where $\mbox{sgn}(J)$ is the sign-function. Thus, depending on the
sign of the coupling constant $J$ both eigenvectors can transform
to the singlet and zeroth component of the triplet. Namely,
\begin{equation}
|\Psi_{1}\rangle=\begin{cases}
|\tau_{0}\rangle=\frac{1}{\sqrt{2}}\left(|\uparrow\downarrow\rangle+|\downarrow\uparrow\rangle\right), & J>0\\
|s\rangle=\frac{1}{\sqrt{2}}\left(|\uparrow\downarrow\rangle-|\downarrow\uparrow\rangle\right), & J<0
\end{cases},\label{Psi1lim}
\end{equation}
and
\begin{equation}
|\Psi_{2}\rangle=\begin{cases}
|s\rangle=\frac{1}{\sqrt{2}}\left(|\uparrow\downarrow\rangle-|\downarrow\uparrow\rangle\right), & J>0\\
|\tau_{0}\rangle=\frac{1}{\sqrt{2}}\left(|\uparrow\downarrow\rangle+|\downarrow\uparrow\rangle\right), & J<0
\end{cases},\label{Psi2lim}
\end{equation}
where we introduce the conventional notations $|\tau_{0}\rangle$
and $|s\rangle$ for triplet and singlet respectively,

The case of $\gamma=0$ is more tricky. First of all, there is no
continuous limit from the eigenvectors $|\Psi_{3,4}\rangle$ to the
upper and lower components of the triplet, $|\tau_{+}\rangle=|\uparrow\uparrow\rangle$
and $|\tau_{-}\rangle=|\downarrow\downarrow\rangle$ as at $\gamma=0$
we have commutativity of the Hamiltonian and $S_{tot}^{z}=S_{1}^{z}+S_{2}^{z}$
operator yielding the block-diagonal form of the Hamiltonian matrix
and decoupling from each other of the basis states, corresponding
to $|\tau_{+}\rangle$ and $|\tau_{-}\rangle$ \cite{oha15}. However,
the eigenvalues admit the corresponding limit leading to
\begin{alignat}{1}
\varepsilon_{3,4}= & \frac{\Delta}{4}\pm\frac{1}{2}|B\left(g_{1}+g_{2}\right)-2K\left(\sigma_{j}+\sigma_{j+1}\right)|,\label{e4lim}
\end{alignat}
depending on the values of the adjacent Ising spins $\sigma_{j}$
and $\sigma_{j+1}$.

In contrast to the case of $|\Psi_{1,2}\rangle$ the result of the
$\gamma=0$ limit depends on the value of the magnetic field.

This condition lead to an obvious critical value of the magnetic field
at which the upper and lower component of the triplet for the vertical
quantum spin dimer are degenerate,
\begin{eqnarray}
B_{c}=\frac{2K(\sigma_{j}+\sigma_{j+1})}{g_{1}+g_{2}}.\label{Bcuplo}
\end{eqnarray}
Namely,
\begin{eqnarray}
|\Psi_{3}\rangle & = & \left\{ \begin{array}{lc}
|\tau_{-}\rangle=|\downarrow\downarrow\rangle, & B>B_{c}\\
|\tau_{+}\rangle=|\uparrow\uparrow\rangle, & B<B_{c},
\end{array}\right.,
\end{eqnarray}
\begin{eqnarray}
|\Psi_{4}\rangle & = & \left\{ \begin{array}{lc}
|\tau_{+}\rangle=|\uparrow\uparrow\rangle, & B>B_{c}\\
|\tau_{-}\rangle=|\downarrow\downarrow\rangle & B<B_{c},
\end{array}\right..
\end{eqnarray}
Here, we use the notation $|\Psi_{3,4}\rangle$ for the $\gamma=0$
having in mind the features mentioned above.

\section{\label{sec:Ground-state-I}On the ground states for Ising diamond chain}

Here we present the ground states and the corresponding energies per the unit cell for the
Ising diamond chain. In our to make the link between Ising and Ising-Heisenberg counterparts more clear and facilitate the interpretation of the Ising limit we keep in general the same notations for the ground states.
\begin{enumerate}
\item Quasi-saturated ($QS_{1}$) state splits into two states

\begin{alignat}{1}
|QS_{1}^{\pm}\rangle= & \prod_{j=1}^{\frac{N}{2}}|\uparrow\rangle_{_{2j-1}}\otimes|\tau_{\pm}\rangle_{_{2j-1}}\otimes|\uparrow\rangle_{_{2j}}\otimes|\tau_{\pm}\rangle_{_{2j}},\\
E_{QS_{1}}^{\pm}= & \frac{\Delta}{2}-\frac{B}{2}(g_{3}+g_{4})\mp\left(B(g_{1}+g_{2})-2K\right),
\end{alignat}

Thus,  $|QS_{1}^{+}\rangle$ corresponds to a saturated (S) state
for $B>B_{c}=\frac{2K}{g_{1}+g_{2}}$, whereas $|QS_{1}^{-}\rangle$
stands for a ferrimagnetic (FIM) state when $B<B_{c}=\frac{2K}{g_{1}+g_{2}}$.
\item Quasi-saturated $QS_{2}$ state splits into two states

\begin{alignat}{1}
|QS_{2}^{\pm}\rangle= & \prod_{j=1}^{\frac{N}{2}}|\downarrow\rangle_{_{2j-1}}\otimes|\tau_{\pm}\rangle_{_{2j-1}}\otimes|\downarrow\rangle_{_{2j}}\otimes|\tau_{\pm}\rangle_{_{2j}},\\
E_{QS_{2}}^{\pm}= & \frac{\Delta}{2}+\frac{B}{2}(g_{3}+g_{4})\mp\left(B(g_{1}+g_{2})+2K\right)
\end{alignat}
Thus, $|QS_{2}^{-}\rangle$ corresponds to a saturated (S) state
for $B<B_{c}=\frac{-2K}{g_{1}+g_{2}}$, whereas  $|QS_{2}^{+}\rangle$
stands for a ferrimagnetic (FIM) state when $B>B_{c}=\frac{-2K}{g_{1}+g_{2}}$.

\item Ferrimagnetic states, $FI_{1}$ and $FI_{2}$

\begin{alignat}{1}
|FI_{1}^{+}\rangle= & \prod_{j=1}^{\frac{N}{2}}|\uparrow\rangle_{_{2j-1}}\otimes|\uparrow\downarrow\rangle_{_{2j-1}}\otimes|\uparrow\rangle_{_{2j}}\otimes|\uparrow\downarrow\rangle_{_{2j}},\\
E_{FI_{1}}^{+}= & -\frac{\Delta}{2}-\frac{B}{2}(g_{3}+g_{4})-B(g_{1}-g_{2})
\end{alignat}

\begin{alignat}{1}\label{B7}
|FI_{2}^{+}\rangle= & \prod_{j=1}^{\frac{N}{2}}|\downarrow\rangle_{_{2j-1}}\otimes|\uparrow\downarrow\rangle_{_{2j-1}}\otimes|\downarrow\rangle_{_{2j}}\otimes|\uparrow\downarrow\rangle_{_{2j}},\\
E_{FI_{2}}^{+}= & -\frac{\Delta}{2}+\frac{B}{2}(g_{3}+g_{4})-B(g_{1}-g_{2})
\end{alignat}

Exchanging the values of the $g_1$ and $g_2$ as well as the orientation of all $S_j$ spins one can get another pair of states, $FI_{1}^{-}$ and $FI_{2}^{-}$ respectively.

\item Antiferromagnetic states, $AF_{1}$ and $AF_{2}$

\begin{alignat}{1}
|AF_{1}^{+}\rangle= & \prod_{j=1}^{\frac{N}{2}}|\uparrow\rangle_{_{2j-1}}\otimes|\uparrow\downarrow\rangle_{_{2j-1}}\otimes|\downarrow\rangle\otimes|\uparrow\downarrow\rangle_{_{2j}}\\
E_{AF_{1}}^{+}= & -\frac{\Delta}{2}-\frac{B}{2}(g_{3}-g_{4})-B(g_{1}-g_{2})
\end{alignat}

\begin{alignat}{1}
|AF_{2}^{+}\rangle= & \prod_{j=1}^{\frac{N}{2}}|\downarrow\rangle_{_{2j-1}}\otimes|\uparrow\downarrow\rangle_{_{2j-1}}\otimes|\uparrow\rangle_{_{2j}}\otimes|\uparrow\downarrow\rangle_{_{2j}}\\
E_{AF_{2}}^{+}= & -\frac{\Delta}{2}+\frac{B}{2}(g_{3}-g_{4})-B(g_{1}-g_{2}).
\end{alignat}

In the full analogy with the previous case, one can also define another pair of the antiferromagnetic states, $AF_{1}^{-}$ and $AF_{2}^{-}$,  by exchanging the values of the Isign spins g-factors simultaneously with the orientation of all $S_j$ spins.

\item "Quantum ferrimagnetic" states $QI_{1}$ and $QI_{2}$

The Ising limit of the Eqs. \eqref{QI1} and \eqref{QI2} lead to the following four ground states of the Ising diamond-chain. Though, some of them have pure antiferromagnetic orientation of spins and other can be described as pure ferrimagnetic ones, we keep the original notations form the Ising-Heisenberg case in order to avoid a confusion.

\begin{alignat}{1}
|QI_{1}^{\pm}\rangle= & \prod_{j=1}^{\frac{N}{2}}|\uparrow\rangle_{_{2j-1}}\otimes|\tau_{\pm}\rangle_{_{2j-1}}\otimes|\downarrow\rangle_{_{2j}}\otimes|\tau_{\pm}\rangle_{_{2j}},\label{B-QI1}\\
E_{QI_{1}}^{\pm}= & \frac{\Delta}{2}-\frac{B}{2}(g_{3}-g_{4})\mp B(g_{1}+g_{2}),
\end{alignat}

\begin{alignat}{1}
|QI_{2}^{\pm}\rangle= & \prod_{j=1}^{\frac{N}{2}}|\downarrow\rangle_{_{2j-1}}\otimes|\tau_{\pm}\rangle_{_{2j-1}}\otimes|\uparrow\rangle_{_{2j}}\otimes|\tau_{\pm}\rangle_{_{2j}},\label{B-QI2}\\
E_{QI_{2}}^{\pm}= & \frac{\Delta}{2}+\frac{B}{2}(g_{3}-g_{4})\mp B(g_{1}+g_{2}).
\end{alignat}

\end{enumerate}

\section{\label{interface-state}Frustrated interface state for Ising diamond
chain}

Here we list possible interfaces between various ground states of the Ising diamond chain with ferromagnetic couplings and mixed, positive and negative, g-factros.

\subsection{One frustrated spin (1/6-fire and 5/6-ice)}

The configuration with one frustrated sublattice ( `1/6-fire and 5/6-ice' in the terms of Ref. [\onlinecite{yin15}]) corresponds to the interface between Quasi-saturated and `Quantum-ferrimagnetic' states when $g_3$ or $g_4$ is negative.

\begin{equation}
B_{c}=\begin{cases}
\frac{2K}{g_{3}}, & QS_{1}^{+}\leftrightarrow QI_{2}^{+},\;\text{and}\;QS_{2}^{+}\leftrightarrow QI_{1}^{+},\\
-\frac{2K}{g_{3}}, & QS_{1}^{-}\leftrightarrow QI_{2}^{-},\;\text{and}\;QS_{2}^{-}\leftrightarrow QI_{1}^{-},\\
\frac{2K}{g_{4}}, & QS_{1}^{+}\leftrightarrow QI_{1}^{+},\;\text{and}\;QS_{2}^{+}\leftrightarrow QI_{2}^{+},\\
-\frac{2K}{g_{4}}, & QS_{1}^{-}\leftrightarrow QI_{1}^{-},\;\text{and}\;QS_{2}^{-}\leftrightarrow QI_{2}^{-}.
\end{cases}
\end{equation}
When spin with $g_{3}$ ($g_{4}$) is frustrated, the
corresponding residual entropy per block is $\mathcal{S}=k_{B}\ln(2)$.

\subsection{Two frustrated spin (1/3-fire and 2/3-ice)}

For the case of two frustrated spins into the unit cell one hase the following interfaces between the ground states of the Ising diamond chain:

interface between `Quantum-ferrimagnetic' and Antiferromagnetic states,
\begin{equation}
B_{c}=\begin{cases}
\frac{\varDelta}{2g_{1}}, & QI_{1}^{+}\leftrightarrow AF_{1}^{-}\;\text{and}\;QI_{2}^{+}\leftrightarrow AF_{2}^{-},\\
-\frac{\varDelta}{2g_{1}}, & QI_{1}^{-}\leftrightarrow AF_{1}^{+}\;\text{and}\;QI_{2}^{-}\leftrightarrow AF_{2}^{+},\\
\frac{\varDelta}{2g_{2}}, & QI_{1}^{+}\leftrightarrow AF_{1}^{+}\;\text{and}\;QI_{2}^{+}\leftrightarrow AF_{2}^{+},\\
-\frac{\varDelta}{2g_{2}}, & QI_{1}^{-}\leftrightarrow AF_{1}^{-}\;\text{and}\;QI_{2}^{-}\leftrightarrow AF_{2}^{-},
\end{cases}
\end{equation}

interface between Quasi-saturated and Ferrimagnetic states,

\begin{equation}
B_{c}=\begin{cases}
\frac{\Delta\pm2K}{2g_{1}}, & QS_{1}^{+}\leftrightarrow FI_{1}^{-}\;\left(QS_{2}^{+}\leftrightarrow FI_{2}^{-}\right),\\
-\frac{\Delta\pm2K}{2g_{1}}, & QS_{2}^{-}\leftrightarrow FI_{2}^{+}\;\left(QS_{1}^{-}\leftrightarrow FI_{1}^{+}\right),\\
\frac{\Delta\pm2K}{2g_{2}}, & QS_{1}^{+}\leftrightarrow FI_{1}^{+}\;\left(QS_{2}^{+}\leftrightarrow FI_{2}^{+}\right),\\
-\frac{\Delta\pm2K}{2g_{2}}, & QS_{2}^{-}\leftrightarrow FI_{2}^{-}\;\left(QS_{1}^{-}\leftrightarrow FI_{1}^{-}\right).
\end{cases}
\end{equation}
There is another interface between the Quasi-Saturated states $QS_{1}^{+}\leftrightarrow QS_{2}^{+}$
and $QS_{1}^{-}\leftrightarrow QS_{2}^{-}$. For this case we have
\begin{equation}
B_{c}=\pm\frac{4K}{g_{3}+g_{4}},
\end{equation}
respectively. Here the spins with $g_{3}$ and $g_{4}$ are
frustrated.

\subsection{Three frustrated spins (1/2-fire and 1/2-ice)}

When three form the six spins in the unit cell are frustrated we one can speak about `1/2-fire and 1/2-ice' configuration. the corresponding interfaces are listed below.

Interface between `Quantum-ferrimagnetic' and Ferrimagnetic ground states, when two
spins with $g_{1}$ ($g_{2}$) and one spin with are disordered
$g_{4}$, the critical magnetic field becomes
\begin{equation}
B_{c}=\begin{cases}
\frac{\Delta}{2g_{1}\pm g_{4}}, & QI_{2}^{+}\leftrightarrow FI_{2}^{-}\;\left(QI_{1}^{+}\leftrightarrow FI_{1}^{-}\right),\\
-\frac{\Delta}{2g_{1}\pm g_{4}}, & QI_{1}^{-}\leftrightarrow FI_{1}^{+}\;\left(QI_{2}^{-}\leftrightarrow FI_{2}^{+}\right),\\
\frac{\Delta}{2g_{2}\pm g_{4}}, & QI_{2}^{+}\leftrightarrow FI_{2}^{+}\;\left(QI_{1}^{+}\leftrightarrow FI_{1}^{+}\right),\\
-\frac{\Delta}{2g_{2}\pm g_{4}}, & QI_{1}^{-}\leftrightarrow FI_{1}^{-}\;\left(QI_{2}^{-}\leftrightarrow FI_{2}^{-}\right),
\end{cases}
\end{equation}
 when two spins with $g_{1}$ (or $g_{2}$) and one
spin with $g_{3}$ are disordered
\begin{equation}
B_{c}=\begin{cases}
\frac{\Delta}{2g_{1}\pm g_{3}}, & QI_{1}^{+}\leftrightarrow FI_{2}^{-}\;\left(QI_{2}^{+}\leftrightarrow FI_{1}^{-}\right),\\
-\frac{\Delta}{2g_{1}\pm g_{3}}, & QI_{2}^{-}\leftrightarrow FI_{1}^{+}\;\left(QI_{1}^{-}\leftrightarrow FI_{2}^{+}\right),\\
\frac{\Delta}{2g_{2}\pm g_{3}}, & QI_{1}^{+}\leftrightarrow FI_{2}^{+}\;\left(QI_{2}^{+}\leftrightarrow FI_{1}^{+}\right),\\
-\frac{\Delta}{2g_{2}\pm g_{3}}, & QI_{2}^{-}\leftrightarrow FI_{1}^{-}\;\left(QI_{1}^{-}\leftrightarrow FI_{2}^{-}\right).
\end{cases}
\end{equation}

There is the interfaces between Quasi-saturated and Antiferromagnetic states.
When two spins with $g_{1}$ (or $g_{2}$) and one
spin with $g_{4}$ are disordered,
\begin{equation}
B_{c}=\begin{cases}
\pm\frac{\Delta+2K}{2g_{1}+g_{4}}, & QS_{1}^{+}\leftrightarrow AF_{1}^{-}\;\left(QS_{2}^{-}\leftrightarrow AF_{2}^{+}\right),\\
\pm\frac{\Delta-2K}{2g_{1}-g_{4}}, & QS_{2}^{+}\leftrightarrow AF_{2}^{-}\;\left(QS_{1}^{-}\leftrightarrow AF_{1}^{+}\right),\\
\pm\frac{\Delta+2K}{2g_{2}+g_{4}}, & QS_{1}^{+}\leftrightarrow AF_{1}^{+}\;\left(QS_{2}^{-}\leftrightarrow AF_{2}^{-}\right),\\
\pm\frac{\Delta-2K}{2g_{2}-g_{4}}, & QS_{2}^{+}\leftrightarrow AF_{2}^{+}\;\left(QS_{1}^{-}\leftrightarrow AF_{1}^{-}\right),
\end{cases}
\end{equation}
When two spins with $g_{1}$ (or $g_{2}$) and one
spin with $g_{3}$ are disordered
\begin{equation}
B_{c}=\begin{cases}
\pm\frac{\Delta+2K}{2g_{1}+g_{3}}, & QS_{1}^{+}\leftrightarrow AF_{2}^{-}\;\left(QS_{2}^{-}\leftrightarrow AF_{1}^{+}\right),\\
\pm\frac{\Delta-2K}{2g_{1}-g_{3}}, & QS_{2}^{+}\leftrightarrow AF_{1}^{-}\;\left(QS_{1}^{-}\leftrightarrow AF_{2}^{+}\right),\\
\pm\frac{\Delta+2K}{2g_{2}+g_{3}}, & QS_{1}^{+}\leftrightarrow AF_{2}^{+}\;\left(QS_{2}^{-}\leftrightarrow AF_{1}^{-}\right),\\
\pm\frac{\Delta-2K}{2g_{2}-g_{3}}, & QS_{1}^{+}\leftrightarrow AF_{2}^{+}\;\left(QS_{1}^{-}\leftrightarrow AF_{2}^{-}\right).
\end{cases}
\end{equation}

\subsection{Four frustrated spins (2/3-fire and 1/3-ice)}

There are several critical points corresponding to four disordered spins in the six-spin unit cell.
The interfaces between two quasi-saturated states ( $QS_{1}^{+}\leftrightarrow QS_{1}^{-}$ and $QS_{2}^{+}\leftrightarrow QS_{2}^{-}$) exist at
\begin{equation}
B_{c}=\pm\frac{2K}{g_{1}+g_{2}},
\end{equation}
respectively. The four spins with the g-factors $g_{1}$ and $g_{2}$ are frustrated here.

Critical magnetic field for the phase boundary between Quasi-ferrimagnetic
and Antiferromagnetic states is given by,
\begin{equation}
B_{c}=\begin{cases}
\frac{\Delta}{2g_{1}\pm(g_{3}-g_{4})}, & QI_{1}^{+}\leftrightarrow AF_{2}^{-}\;\left(QI_{2}^{+}\leftrightarrow AF_{1}^{-}\right),\\
\frac{-\Delta}{2g_{1}\pm(g_{3}-g_{4})}, & QI_{2}^{-}\leftrightarrow AF_{1}^{+}\;\left(QI_{1}^{-}\leftrightarrow AF_{2}^{+}\right),\\
\frac{\Delta}{2g_{2}\pm(g_{3}-g_{4})}, & QI_{1}^{+}\leftrightarrow AF_{2}^{+}\;\left(QI_{2}^{+}\leftrightarrow AF_{1}^{+}\right),\\
\frac{-\Delta}{2g_{2}\pm(g_{3}-g_{4})}, & QI_{2}^{-}\leftrightarrow AF_{1}^{-}\;\left(QI_{1}^{-}\leftrightarrow AF_{2}^{-}\right).
\end{cases}
\end{equation}
Whereas the interface between Quasi-ferrimagnetic and Ferrimagnetic phases by
\begin{equation}
B_{c}=\begin{cases}
\frac{\pm(\Delta+2K)}{2g_{1}+(g_{3}+g_{4})}, & QS_{1}^{+}\leftrightarrow FI_{2}^{-}\;\left(QS_{2}^{-}\leftrightarrow FI_{1}^{+}\right),\\
\frac{\pm(\Delta-2K)}{2g_{1}-(g_{3}+g_{4})}, & QS_{2}^{+}\leftrightarrow FI_{1}^{-}\;\left(QS_{1}^{-}\leftrightarrow FI_{2}^{+}\right),\\
\frac{\pm(\Delta+2K)}{2g_{2}+(g_{3}+g_{4})}, & QS_{1}^{+}\leftrightarrow FI_{2}^{+}\;\left(QS_{2}^{-}\leftrightarrow FI_{1}^{-}\right),\\
\frac{\pm(\Delta-2K)}{2g_{2}-(g_{3}+g_{4})}, & QS_{2}^{+}\leftrightarrow FI_{1}^{+}\;\left(QS_{1}^{-}\leftrightarrow FI_{2}^{-}\right).
\end{cases}
\end{equation}
In all these sixteen cases the spins with $g_{1}$ (or $g_{2}$) and one both spins with $g_{3}$ and $g_{4}$ are frustrated.

\subsection{Five frustrated spins (5/6-fire and 1/6-ice)}

Finally, the spin configuration with five frustrated spins in the six-spin unit cell is possible at the interface between quasi-saturated and "quasi-ferrimagnetic" states.  The phase boundary of these `5/6-fire and 1/6-ice' sates are given by

\begin{equation}
B_{c}=\begin{cases}
\frac{2K}{2(g_{1}+g_{2})\pm g_{4}}, & QS_{1}^{+}\leftrightarrow QI_{1}^{-}\;\left(QS_{1}^{-}\leftrightarrow QI_{1}^{+}\right),\\
\frac{-2K}{2(g_{1}+g_{2})\pm g_{4}}, & QS_{2}^{-}\leftrightarrow QI_{2}^{+}\;\left(QS_{2}^{+}\leftrightarrow QI_{2}^{-}\right),\\
\frac{2K}{2(g_{1}+g_{2})\pm g_{3}}, & QS_{1}^{+}\leftrightarrow QI_{2}^{-}\;\left(QS_{1}^{-}\leftrightarrow QI_{2}^{+}\right),\\
\frac{-2K}{2(g_{1}+g_{2})\pm g_{3}}, & QS_{2}^{-}\leftrightarrow QI_{1}^{+}\;\left(QS_{2}^{+}\leftrightarrow QI_{1}^{-}\right).
\end{cases}
\end{equation}
It is easy to recognizing looking at the denominators that in all cases two spins
with $g_{1}$ g-factor, two spins with $g_{2}$, one of the spins with $g_{3}$
and $g_{4}$ are frustrated.

\end{document}